\documentclass[aps,pre,reprint,amsmath,superscriptaddress]{revtex4-1}
\usepackage{subfigure}
\usepackage{graphicx}
\usepackage{bm}
\usepackage{natbib}
\usepackage{amssymb}
\usepackage{amsbsy}
\usepackage{color}
\usepackage{mathrsfs,mathcomp}

\begin{document}

\title{How water droplets evaporate on a superhydrophobic substrate}

\author{Hanneke Gelderblom}
\author{\'Alvaro G. Mar\'in}
\affiliation{Physics of Fluids, Faculty of Science and Technology, University of Twente, P.O. Box 217, 7500 AE Enschede, The Netherlands}
\author{Hrudya Nair}
\author{Arie van Houselt}
\author{Leon Lefferts}
\affiliation{Catalytic Processes and Materials, Faculty of Science and Technology,  P.O. Box 217, 7500 AE Enschede,University of Twente}
\author{Jacco H. Snoeijer}
\author{Detlef Lohse}
\affiliation{Physics of Fluids, Faculty of Science and Technology, University of Twente, P.O. Box 217, 7500 AE Enschede, The Netherlands}

\date{\today}

\begin{abstract}
Evaporation of water droplets on a superhydrophobic substrate, on which
the contact line is pinned, is investigated. While previous studies mainly focused on droplets with contact angles smaller than $90^\circ$, here we analyze almost the full range of possible contact angles (10$^\circ$-150$^\circ$). The greater contact angles and pinned contact lines can be achieved by the use of superhydrophobic Carbon Nanofiber substrates. The time-evolutions of the contact angle and the droplet mass are examined. 
The experimental data is in good quantitative agreement with the model presented by Popov (Physical Review E 71, 2005), demonstrating that the evaporation process is quasi-static,  diffusion-driven, and that thermal effects play no role. Furthermore, we show that the experimental data for the evolution of both the contact angle and the droplet mass can be collapsed onto one respective universal curve for all droplet sizes and initial contact angles.

\end{abstract}

\maketitle
\section{Introduction}

Evaporation of sessile droplets with small contact angles ($<90^\circ$) has been studied extensively. Several evaporation modes have been explored: the constant contact-angle mode \cite{Picknett:1977p108, Erbil:2002p1786}, in which the contact area of the droplet on the substrate vanishes; the constant contact-area mode \cite{Picknett:1977p108, Birdi:1989p8, ROWAN:1995p459, Deegan:2000p11, Hu:2002p192}, in which the contact angle vanishes; and the combination of both modes \cite{Picknett:1977p108, BourgesMonnier:1995p10, Cachile:2002p413}. A thorough understanding of droplet evaporation is of vital importance for examining the drying rate \cite{Picknett:1977p108, Birdi:1989p8, BourgesMonnier:1995p10, Hu:2002p192,David:2007p15, Dunn:2008p62, Dunn:2009p3}, the flow patterns observed inside drying drops \cite{Hu:2005p522,Hu:2005p523, Ristenpart:2007p90}, and the residual deposits \cite{Deegan:1997p12, Deegan:2000p11,Popov:2005p16}.  

In early modeling of evaporating drops \cite{Birdi:1989p8, BourgesMonnier:1995p10, ROWAN:1995p459, Erbil:1997p2183}, the evaporative flux was assumed to be uniform in the radial direction, as it is for evaporation from a sphere. However, in his study of contact-line deposits, Deegan \cite{Deegan:2000p11} argued that the evaporative flux from a sessile drop with a spherical cap shape is generally not uniform, but diverges near the edge of the drop for contact angles smaller than $90^\circ$.  Hu and Larson \cite{Hu:2002p192}  later used a numerical model to find an expression for the rate of mass loss from a drop in terms of its contact angle, taking this divergence into account. Their model applies to contact angles smaller than $90^\circ$.

For larger contact angles, few theoretical descriptions exist for diffusion around a spherical-cap droplet.
In \cite{Picknett:1977p108}, the rate of mass loss was expressed in terms of a series solution, which can be approximated in both the small and the large contact angle regimes. Popov \cite{Popov:2005p16} described an analytical solution for the rate of mass loss in terms of the contact angle, which applies to the full range of contact angles. However, this model has never been validated against experimental data in the large contact-angle regime.

Apart from the diffusive spreading of water vapor described by the models mentioned above, there are other factors that may influence the evaporation rate (see e.g. \cite{Cazabat:2010p1660} for an overview). Firstly, the evaporation models discussed assume a stationary contact line. When the contact line is moving, dynamic effects may complicate the problem, for both the vapor concentration outside and the viscous flow inside the drop \cite{Cachile:2002p413}.
Secondly, evaporative cooling of the drop can reduce the evaporation rate \cite{David:2007p15, Dunn:2008p62, Dunn:2009p3}. The resulting temperature gradients on the drop surface can induce a Marangoni flow \cite{Hu:2005p523, Ristenpart:2007p90}, and can give rise to a Marangoni-B\'enard instability \cite{Nguyen:2002p2373}. Finally, in addition to the diffusion of water vapor, free convective transport may play a role, increasing the evaporation rate \cite{Dunn:2009p3, ShahidzadehBonn:2006p409}. However, the influence of these factors on the evaporation rate still has to be confirmed experimentally.
\begin{figure}
	\includegraphics[width=3.4 in]{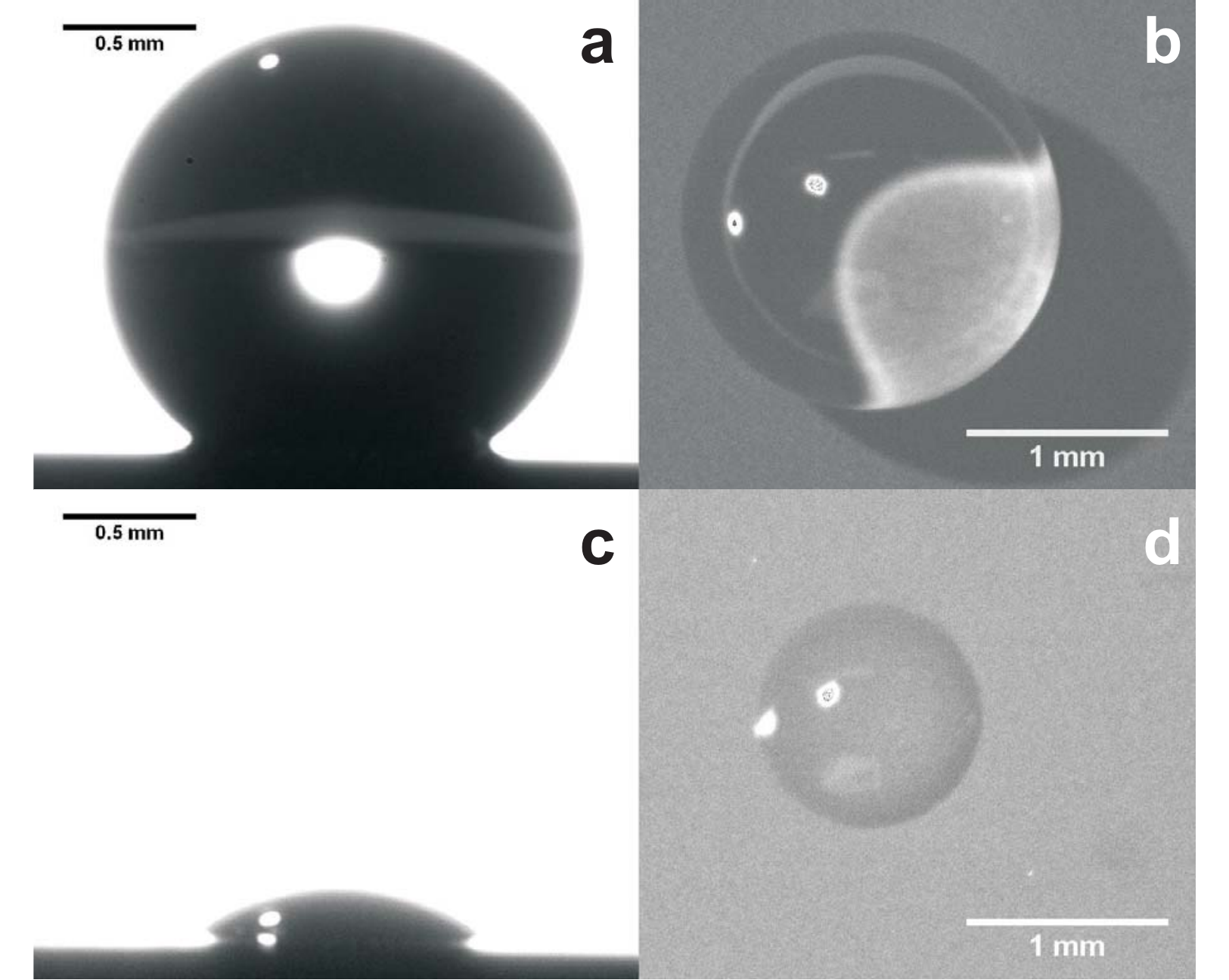}
	\caption{Side view (a) and top view (b) of a 8 $\mu$l droplet on a CNF substrate in the initial moments. (c) and (d): the same droplet in the last moments before being completely evaporated. Note that the contact line remains perfectly circular and completely pinned until almost the end of the process. \label{drops}}
\end{figure}

In this paper, we describe our investigation into the evaporation of water droplets on Carbon Nanofiber (CNF) substrates; see Fig. \ref{drops}. These substrates belong to the family of ordered carbonaceous structures: the graphitic planes are oriented under an angle to the central axis. CNF substrates can exhibit superhydrophobicity \cite{Tsai:2009p2189}. The samples used here have contact angles with water ranging from 150$^\circ$ up to 170$^\circ$.
On superhydrophobic substrates, all evaporation modes can occur; the constant contact-angle mode is mostly observed when the contact angle hysteresis is low, the constant contact-area mode when the hysteresis is high \cite{Kulinich:2009p1556, Zhang:2006p1551}. On our CNF substrates, the contact line remains pinned throughout nearly the entire experiment, hence evaporation takes place in the constant contact-area mode. By contrast, superhydrophobic substrates based on micropillar arrays display contact-line jumps during evaporation \cite{Sbragaglia:2007p2212, Tsai:2010p2188}. Since we consider pinned contact lines, we can study evaporating drops in almost the full range of possible contact angles (0$^\circ$-150$^\circ$). The rate of mass loss and contact-angle evolution over time are obtained experimentally for various drop sizes. We show that the evaporation dynamics is accurately described by the diffusion-based model of Popov \cite{Popov:2005p16}, suggesting that thermal and free-convection effects are unimportant in our experiment. In addition, we show that the evolutions of the droplet mass and contact angle can be described by a universal relation, that is, independent of the drop size and initial contact angle.

In Sec. \ref{experiments}, the experimental set-up and preparation of the CNF substrates are described. The experimental results are discussed in Sec. \ref{exres}. The theoretical model for droplet evaporation adopted from Popov \cite{Popov:2005p16} is briefly described in Sec. \ref{theory}. Finally, in Sec. \ref{comp}, it is shown that the theoretical results are in good quantitative agreement with the experimental data.

\section{Experimental methods}\label{experiments}
\subsection{Preparation of the CNF substrates}
\begin{figure}
\includegraphics[width=3.4 in]{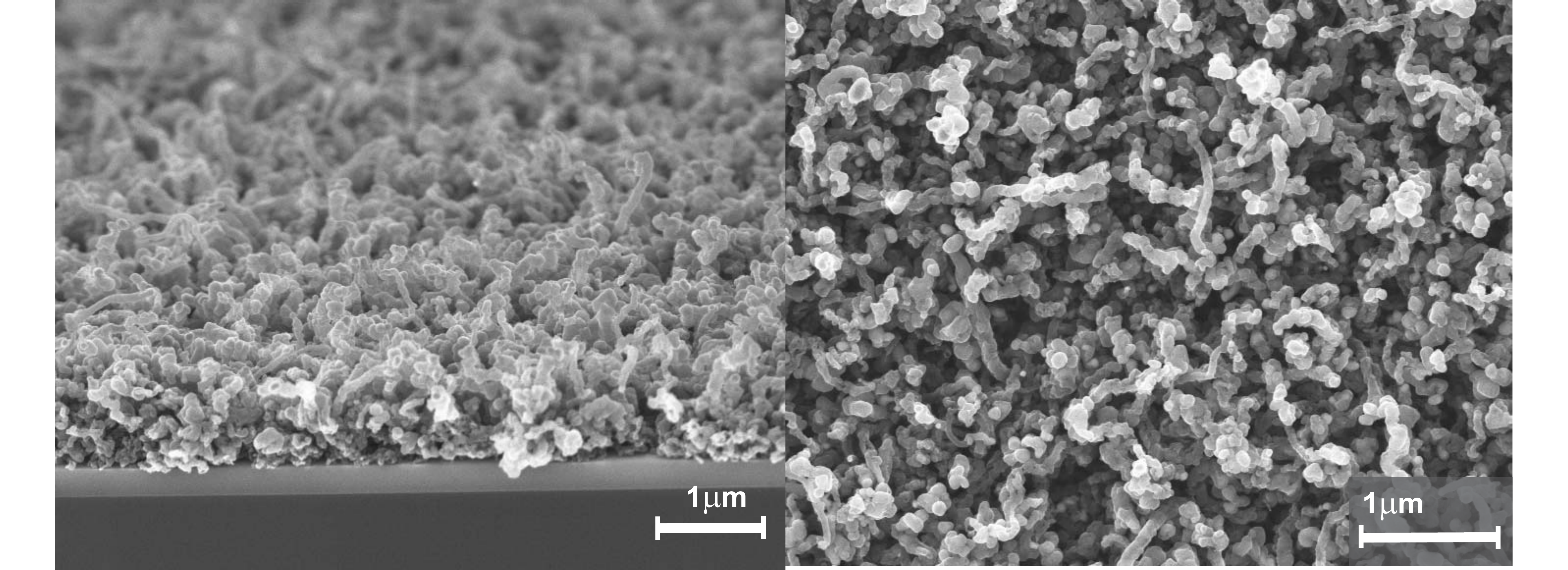}
\caption{SEM images of the Carbon Nanofibers (CNFs) used as superhydrophobic substrates. Tilted side view (left) and augmented top view (right).\label{CNFs}}
\end{figure}
The droplets were left evaporating in an empty room \footnote{No human heat sources were present.} at a constant temperature of $23\tccentigrade$ and a humidity of 30\% over Carbon Nanofiber substrates; see Fig. \ref{CNFs}. Carbon nanofibers were grown on oxidized silicon substrates using a Ni thin film as catalyst. A 250 nm thick SiO$_{2}$ layer was grown on \textit{p}-type Si(001) via wet oxidation. On top of this oxide layer 10 nm Ta was deposited, followed by a 25 nm thick Ni layer. The samples were pretreated prior to the CNF synthesis in a quartz reactor. The substrates were placed on a flat quartz boat placed centrally inside a quartz reactor, and the temperature was increased at a rate of 5$^\circ$C min$^{-1}$ from room temperature up to 500$^\circ$C in a N$_{2}$ (99.999$\%$, Indugas) atmosphere. During this pretreatment step, the samples were subjected to 20 vol.$\%$ of H$_{2}$ in N$_{2}$ at a total flow rate of 50 ml min$^{-1}$ at 500  $^\circ$C for 2 hours; then the temperature was increased up to 635$^\circ$C. At 635 $^\circ$C, 25 vol.$\%$ ethylene (99.95$\%$ Praxair) in N$_{2}$ was passed through the reactor for 1 hour, while 6.25 vol.$\%$ H$_{2}$ (99.999$\%$, Indugas) was added for the first minutes of the reaction time. After the reaction time, the substrates were cooled down in N$_{2}$ at a rate of 10$^\circ$C min$^{-1}$ until room temperature was reached. The CNF samples were used without further functionalization.

\subsection{Measurement of droplet evaporation}
To analyze the evaporation of droplets on CNF substrates, the droplets were observed during their total evaporation time and photographed at 1 s time intervals. Two synchronized cameras (Lumenera Lm135, 1392 x 1040 pixels) were used for this purpose, one taking side-view images and another one taking 
top-view images; see Fig. \ref{drops}. Side-view images allowed us to compute volume (mass), contact angle, area, droplet radius, mass loss, and spreading velocity at every instant. The image analysis was performed using a custom-made MATLAB code in which the detected droplet profile was fitted to an ellipse. The droplets considered in this study are much smaller than the capillary length (which is 2.7 mm for a water droplet \cite{Cazabat:2010p1660}), hence we can neglect flattening of the drops by gravity. Nevertheless, we used an elliptical rather than a spherical fitting. The elliptical fit allowed us to use three fitting parameters (two semi axes and the angle of the ellipse with the horizontal plane) instead of only one (droplet radius), thereby increasing the precision of the determination of the volume and contact angle of the droplets. The ellipticity of the droplets, defined as the ratio between both semi axes, was always below 7\%.

The contact line of the droplets was detected automatically; the contact angles were then measured by finding the tangent of the ellipse at the contact line. The error in the determination of the contact angle, based on the quality of the fits, was found to be below 1\%. The volume of the droplet was obtained by calculating the ellipse area above the contact line and assuming rotational symmetry with respect to the vertical axis, with an error below 10\%. The rate of mass loss was computed applying a fourth-order finite differentiation of the ellipse volume over time.

Top-view images were used to obtain qualitative information on the stability and circular symmetry of the contact line; using this information, we rejected those few experiments in which the contact line had a highly irregular shape.

Due to the chaotic three-dimensional distribution of the nanofibers, the way the liquid wets the structure is more complex than for ordered superhydrophobic microstructures \cite{Sbragaglia:2007p2212, Tsai:2010p2188}, for which two wetting states can be defined: the \textit{Cassie-Baxter state}, in which the contact of the liquid with the substrate is minimum, and the \textit{Wenzel state}, in which the contact is maximum. In our case, it is assumed that the liquid remains in a mixed state and that the transitions from one intermediate state to another are sufficiently smooth to be undetectable. Therefore, we will not use this terminology in this paper.

\section{Experimental results}\label{exres}

\begin{figure}
	\subfigure[]{
	\includegraphics[width=3.4 in]{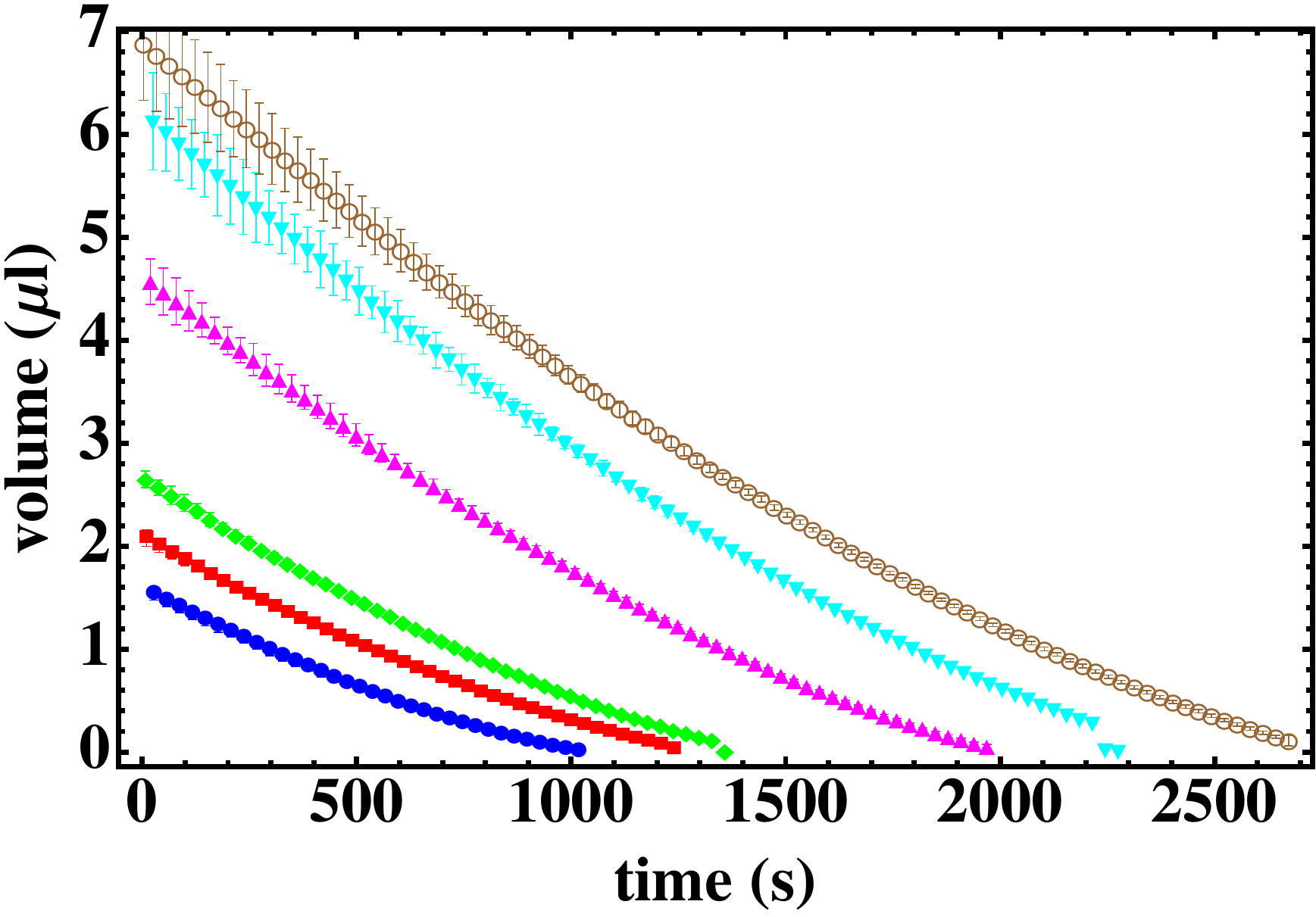}}
	\subfigure[]{
 \includegraphics[width=3.4 in]{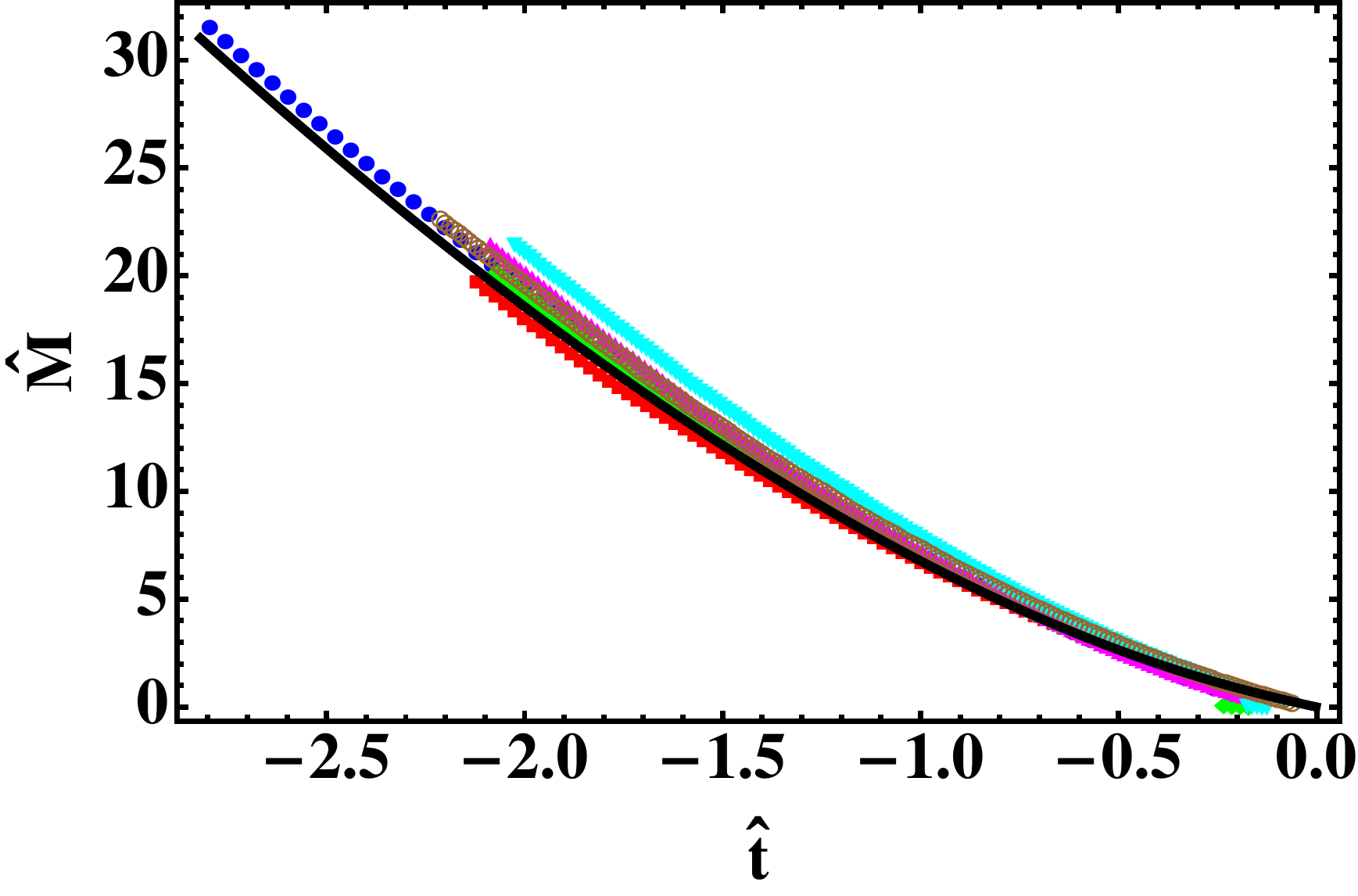}}
  \caption{\label{vol} (Color online) (a) Droplet volume versus time for initial droplet volumes of 1.6 $\mu$l (blue filled circles), 2.1 $\mu$l (red squares), 2.9 $\mu$l (green diamonds), 4.6 $\mu$l (magenta upward triangles), 6.2 $\mu$l (cyan downward triangles), and 6.9 $\mu$l (brown unfilled circles). The error bars are deduced from the elliptical fit to the data. The measurements were performed with a time resolution of 1 s, but for clarity we show the data with a 30 s resolution. (b) The dimensionless droplet mass plotted against the dimensionless time. The black solid line represents the theoretical prediction according to the Popov model. The experimental data is scaled according to (\ref{dimless}). The time is set to 0 at the end of the droplet life (see text).}
\end{figure}

\begin{figure}
	\subfigure[]{
	\includegraphics[width=3.4 in]{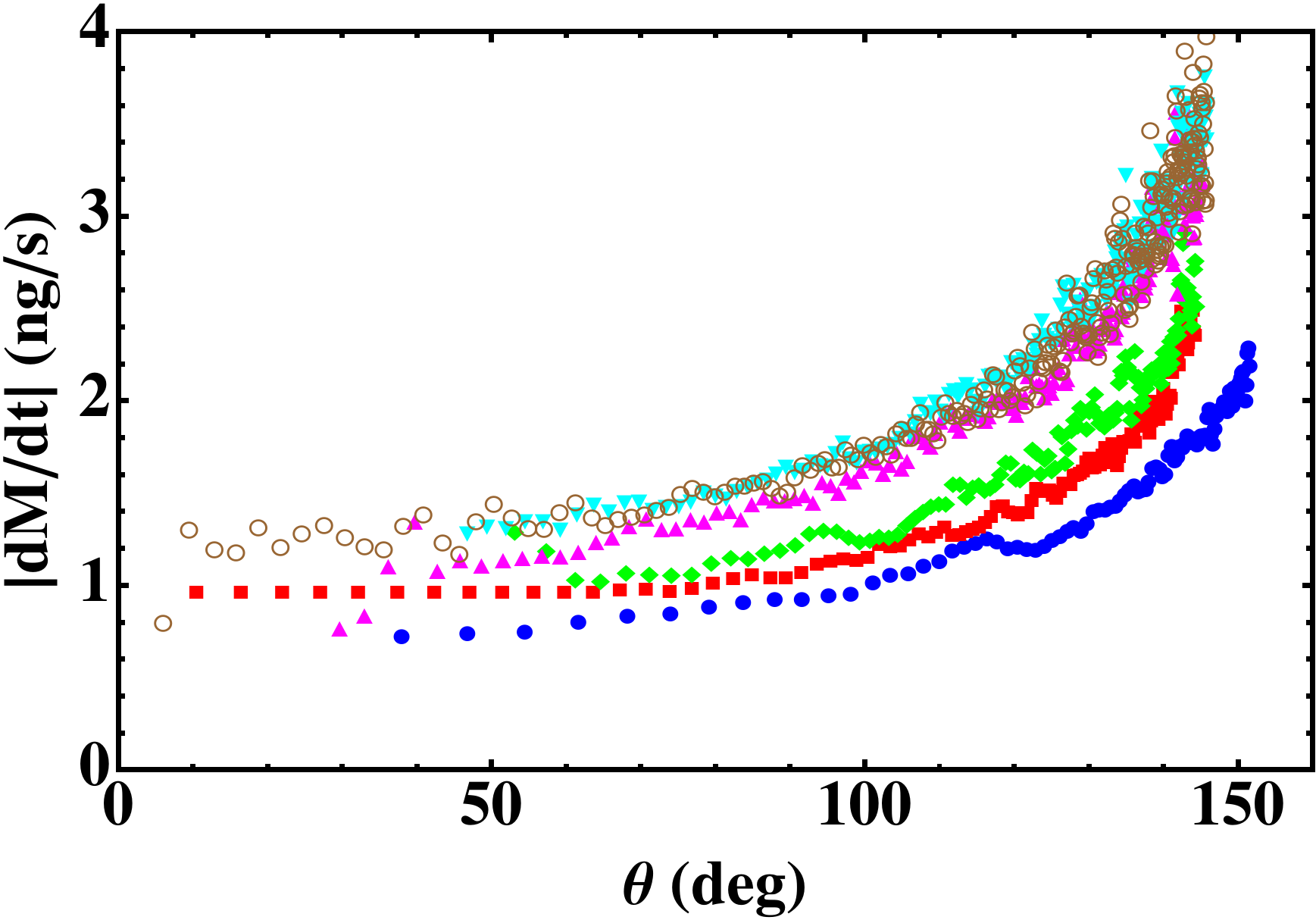}}
	\subfigure[]{
 \includegraphics[width=3.4 in]{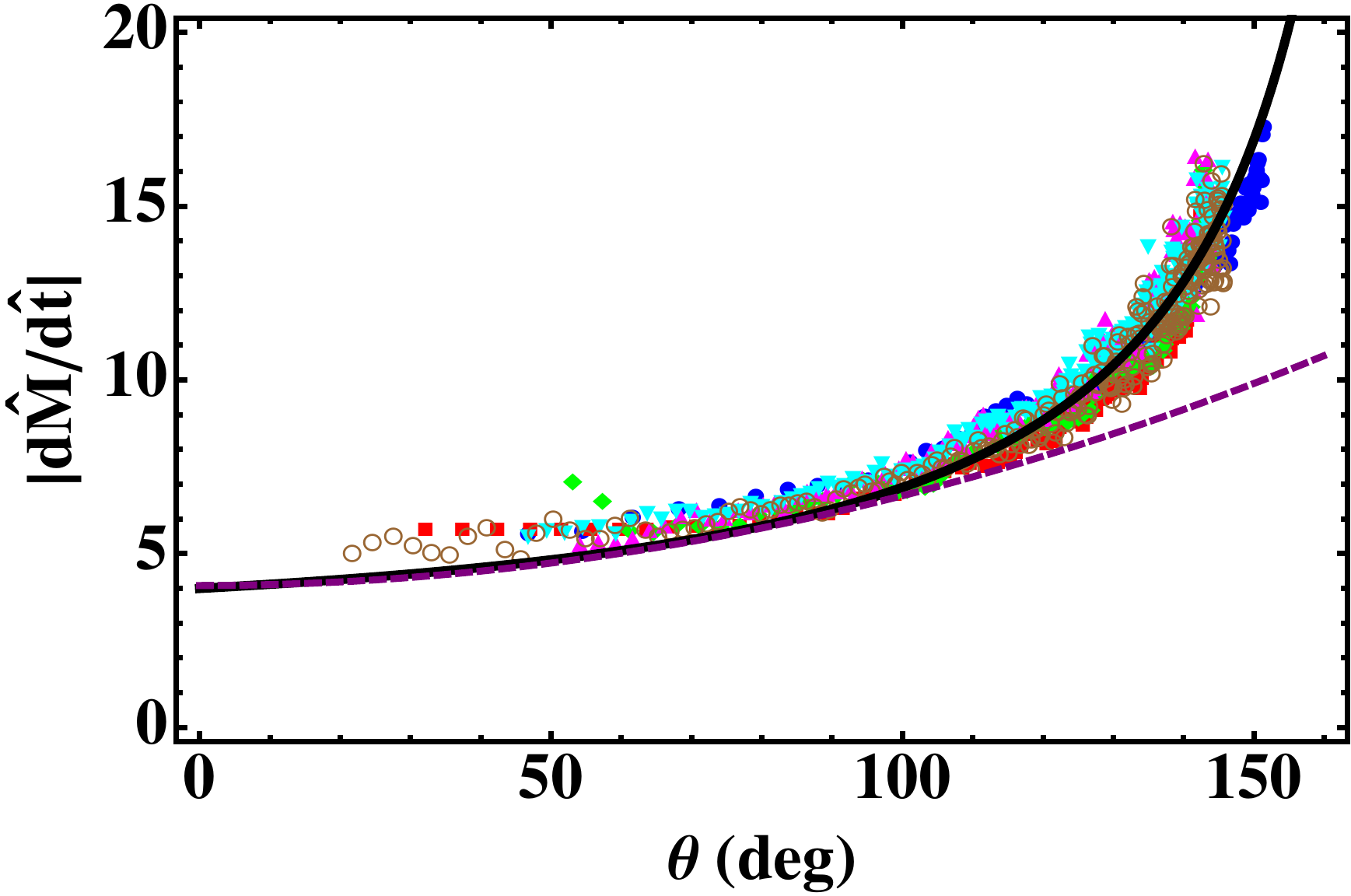}}
	\caption{\label{dmdt} (Color online) (a) The rate of mass loss of the droplet (derived from the measured droplet volume) versus the contact angle. Colors and markers are as in Fig. \ref{vol}. (b) The same data, but now scaled according to (\ref{dimless}). Predictions from the Popov model (black solid line) and the model of Hu \& Larson (purple dashed line) are shown.}
\end{figure}

The droplet volume, contact angle, and radius were determined from the experimental data with a time resolution of 1 s. The droplet volume versus time plot, clearly shows nonlinear behavior; see Fig. \ref{vol}(a). Hence, a model based on the small contact-angle approximation, which predicts the droplet volume to decrease linearly in time \cite{Deegan:2000p11, Hu:2002p192}, will not suffice to describe the evolution of the droplet volume over time.
From the droplet volume measurements, the rate of mass loss of the droplet $dM/dt$ was derived, as described in Sec. \ref{experiments}. Figure \ref{dmdt}(a) shows that $dM/dt$ decreases with decreasing contact angle; hence it also decreases in time. Again, nonlinear behavior is observed, with a steep decline for larger contact angles, but this levels off to a constant rate of mass loss for contact angles smaller than 70$^{\circ}$.

\begin{figure}
\subfigure[]{
	\includegraphics[width=3.4 in]{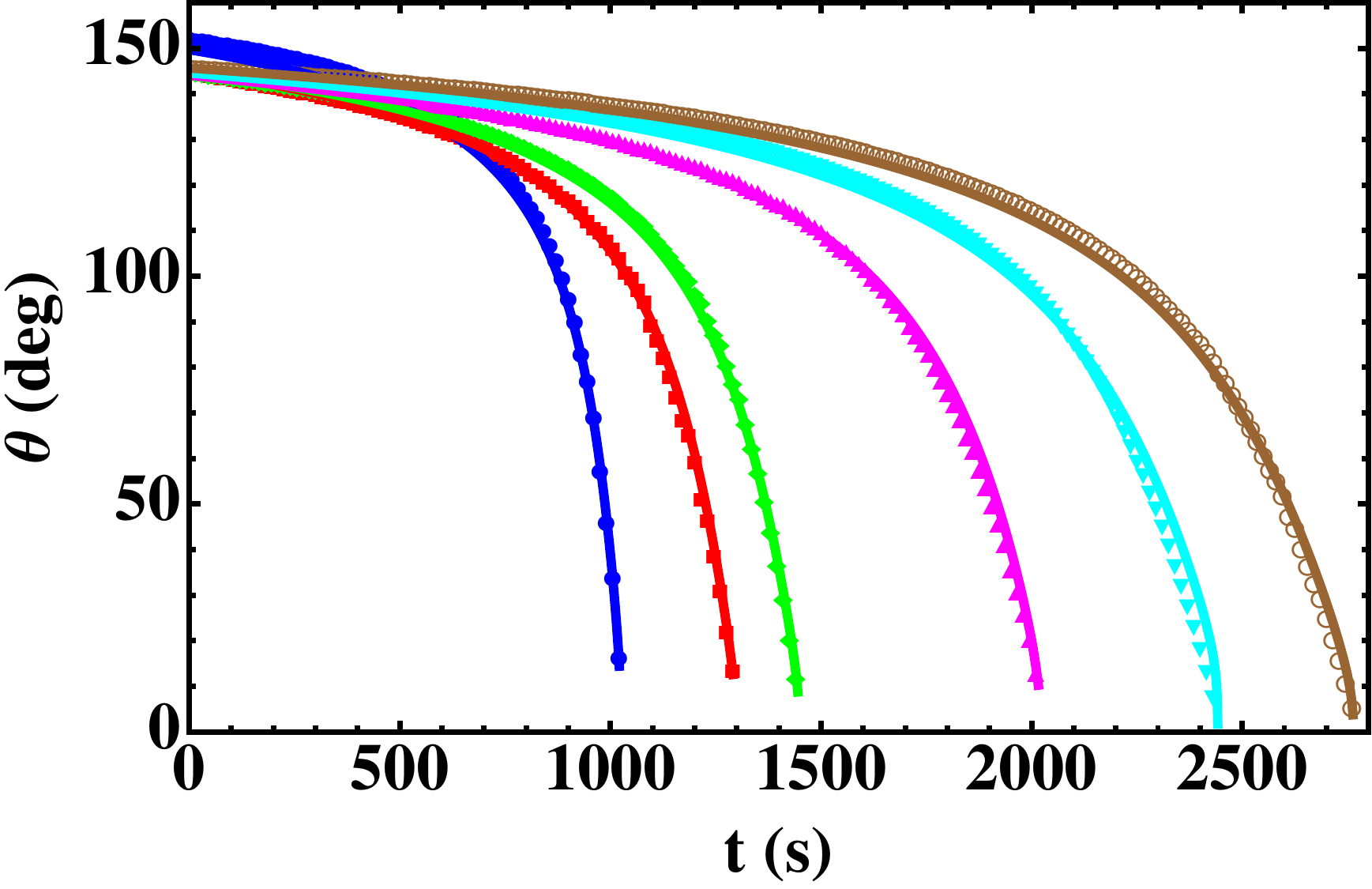}}
	\subfigure[]{
 \includegraphics[width=3.4 in]{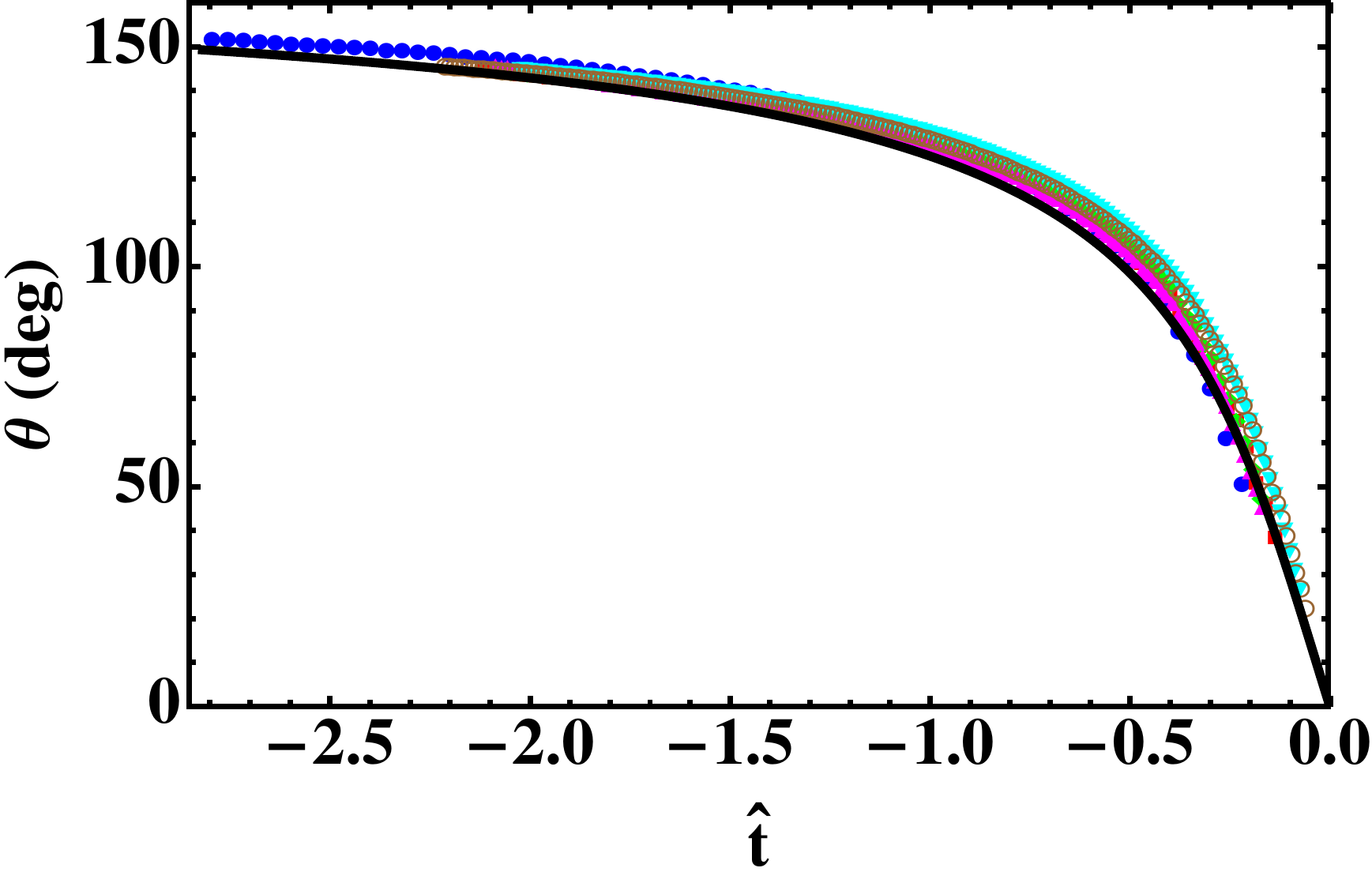}}
	\caption{\label{modex} (Color online) (a) The evolution of the contact angle over time. The experimental data ($\bullet$) can be very well described by the theoretical model of Popov (---), by adjusting the drop radius according to its experimental value (see Sec. \ref{comp}). The error in the experimental data is not shown, since it is below 1$\%$. (b) The same data, but with the time scaled according to (\ref{dimless}), and set to 0 at the end of the droplet life (see text). The black solid line represents the theoretical prediction according to the Popov model. Colors and markers are as in Fig. \ref{vol}.}
\end{figure}

During the evaporation, the contact angle of the droplets decreases over time from about 150$^\circ$ to 0$^\circ$, as shown in Fig. \ref{modex}(a). Initially, the contact angle decreases slowly over time. This is followed by a more rapid, linear decrease over time when the contact angle becomes smaller than approximately 70$^{\circ}$. The initial contact angles of the droplets differ somewhat owing to irregularities in the substrate. For comparison, not only the experimental data, but also the predictions based on the Popov model are shown in Fig. \ref{modex}(a). A more detailed explanation of this model is given in Sec. \ref{theory}.

\begin{figure}
	\includegraphics[width=3.4 in]{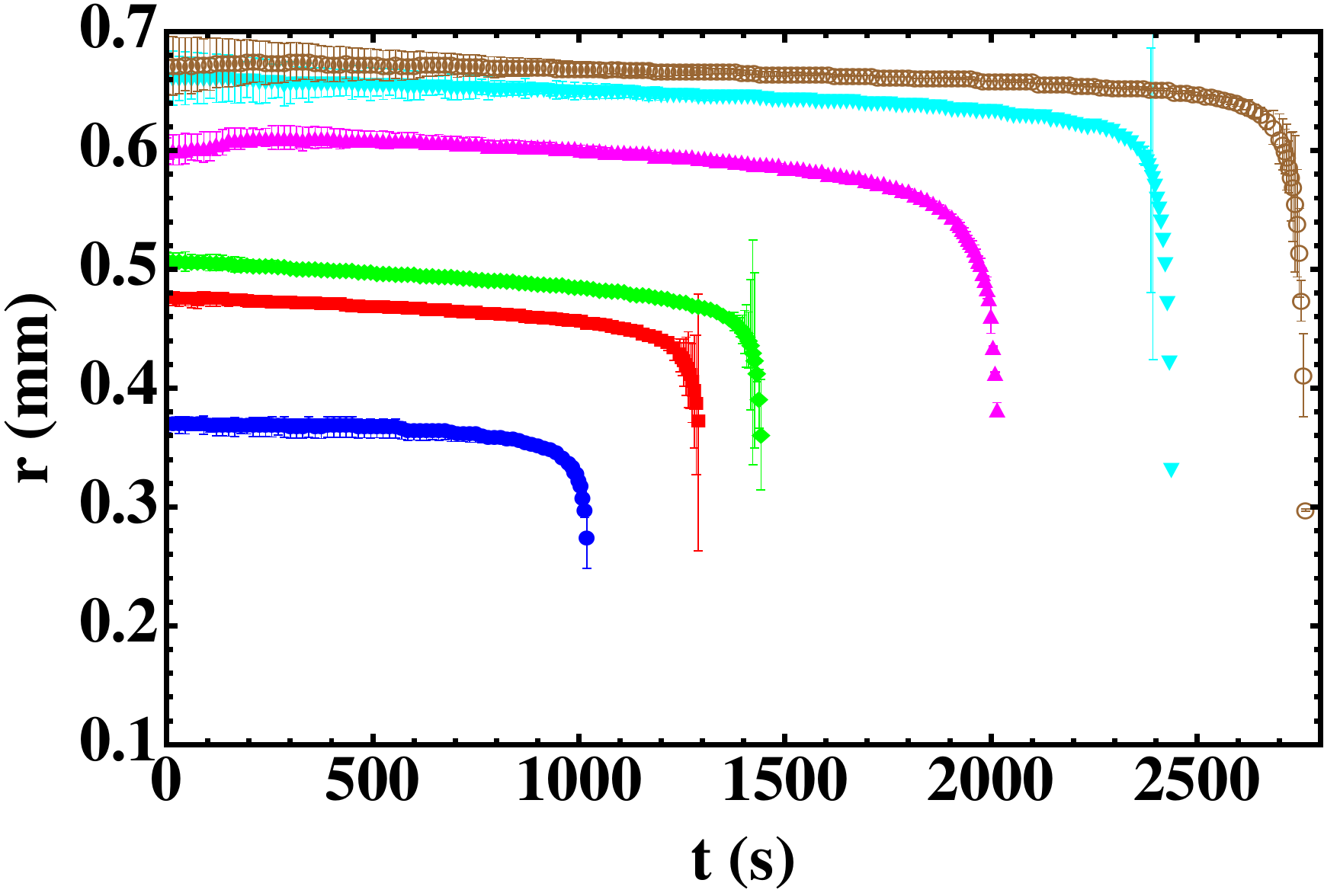}
	\caption{\label{rad} (Color online) The droplet radius versus time. Significant depinning of the contact line is observed during the final 4\% of the droplet lifetime. Data are shown with 15 s time-resolution. During the depinning, a resolution of 5 s is used. Colors and markers are as in Fig. \ref{vol}.}
\end{figure}

One advantage of the CNF substrates is that the contact lines of the droplets remain pinned throughout almost the entire experiment. Therefore, droplet evaporation in the constant contact-area mode can be studied, in the absence of any contact-line dynamics. Similar behavior of the contact angle in the pinned situation has been reported for natural lotus leaves \cite{Zhang:2006p1551}, synthetic superhydrophobic surfaces with high contact angle hysteresis \cite{Kulinich:2009p1556}, and aligned carbon nanotube (CNT) samples \cite{Gjerde:2008p2356}. Figure \ref{rad} shows that depinning only occurs during the final moments of the droplet life. In the depinning phase, the contact angle is typically smaller than 40$^{\circ}$. Once the droplet starts to depin, also the measurements error shoots up, since the contact line does not depin homogeneously, and is therefore no longer exactly circular.

\section{Theory of droplet evaporation}\label{theory}
To describe theoretically the measured time-evolution of a droplet's contact angle and mass, we need to know the evaporative flux from the droplet surface. This flux depends on the rate-limiting step in the vapor transport. We assume that vapor transport by free convection, induced by the density difference between dry and humid air \cite{ShahidzadehBonn:2006p409}, is negligible compared to diffusive transport. The influence of evaporative cooling of the droplet on the evaporation rate \cite{Dunn:2009p3} is also neglected. Hence, the vapor transport occurs mainly by diffusive spreading of the water vapor in air, and is characterized by diffusion time $t_d=R^2/D$, with $R$ the droplet radius in the plane of the substrate, and $D$ the diffusion coefficient. The diffusion time for water vapor in air is of the order of $10^{-2}$ s. The evaporation occurs in a quasi-steady fashion: the timescale for diffusion is much smaller than the typical droplet evaporation time $t_e$. As will become clear from the dimensional analysis presented in (\ref{dimless}), $t_e=\rho/(c_s-c_{\infty})~t_d$. In essence, $t_e$ can be estimated by comparing the initial droplet mass, proportional to the droplet density $\rho$, to the rate of mass loss, proportional to $c_s-c_{\infty}$, the vapor concentration difference between the drop surface and the surroundings. Here, $t_e/t_d=\rho/(c_s-c_{\infty})$ is of the order of $10^{5}$.
We do not take into account the Kelvin correction to the vapor pressure, since this effect is negligible for droplets of the size considered here.

To determine the diffusive outflux from the drop surface, the vapor concentration field around the droplet has to be calculated. We follow the approach taken by Popov \cite{Popov:2005p16}. For completeness, we briefly formulate the problem below. 

A cylindrical coordinate system ($r, z, \phi$) is adopted, with $r$ being the radial coordinate, $z$ the direction normal to the substrate, and $\phi$ the circumferential coordinate. The origin of this system is chosen such that $z=0$ corresponds to the substrate, and $r=0$ to the center of the droplet. In this case the problem is axisymmetric, i.e. $\phi$-independent.  In the quasi-steady, diffusion-limited case the concentration field $c(r,z)$ around the droplet is given by:
\begin{equation}\label{laplace}
\nabla^2 c=0.
\end{equation}
The boundary conditions imposed along the spherical-cap shaped droplet with arbitrary contact angle $\theta$ are: (i) $c=c_s$, the saturated vapor concentration, along the droplet surface; (ii) $c=c_{\infty}$, the ambient vapor concentration, far away from the drop; and (iii) the substrate is impermeable, hence $\partial c/\partial z=0$ along the substrate. The diffusive flux is given by $\bm{J}=-D\nabla c$. In our experiments, the ambient temperature was $23^\circ$C, the humidity $H=0.3$. At this temperature, $D=24.6\times10^{-6}$ m$^2$/s, $\rho=997.6$ kg/m$^3$, and $c_s=2.08 \times 10^{-2}$ kg/m$^3$ (obtained from \cite[p. 6-1, 6-191]{CRC} by linear interpolation); furthermore $c_\infty=Hc_s$.

In the limit of small contact angles, simplified solutions to (\ref{laplace}) subject to the boundary conditions (i)-(iii) can be used, as presented by Deegan \cite{Deegan:2000p11} and Hu \& Larson \cite{Hu:2002p192}. In our case a more advanced model is needed, since we consider droplets evaporating on a superhydrophobic substrate, with initial contact angles of approximately 150$^{\circ}$. The analytical solution to the equivalent problem of finding the electric potential around a charged lens-shaped conductor is described in \cite{Lebedev1965}. Popov \cite{Popov:2005p16} used this result to determine the rate of mass loss from a droplet of arbitrary contact angle:
\begin{eqnarray}\label{dmdttheory}
\frac{dM}{dt}&=&-\int_0^R J(r)\sqrt{1+(\partial_rh)^2}2\pi r dr\\ \nonumber
&=&-\pi RD(c_s-c_\infty)f(\theta),
\end{eqnarray}
with $M$ the droplet mass, $J$ the diffusive outflux from the droplet surface, $h(r,t)$ the droplet height, $t$ the time, and
\begin{eqnarray}\label{f}
f(\theta)=&&\\ \nonumber
&&\frac{\sin \theta}{1+\cos \theta}+4\int_0^\infty \frac{1+\cosh 2\theta\tau}{\sinh 2\pi\tau}\tanh[(\pi-\theta)\tau]d\tau.
\end{eqnarray}
The droplet mass can be expressed in terms of $\theta$ by the geometric relation
\begin{equation}
	M(\theta)=\rho \pi R^3\frac{\cos^3\theta-3\cos\theta+2}{3\sin^3\theta},\label{mfromth}
\end{equation}
which yields an ordinary differential equation for $\theta$ as a function of $t$
\begin{eqnarray}
\frac{d\theta}{dt}=-\frac{D(c_s-c_\infty)}{\rho R^2}(1+\cos\theta)^2f(\theta).\label{thetatheory}
\end{eqnarray}
Numerical integration then gives  $\theta$ as a function of $t$. Once $\theta$ is known, $M(\theta)$ and $dM/dt$ can be derived.

In Fig. \ref{vol}(a)-\ref{modex}(a) we showed the evolution of the droplet mass and contact angle in time for various drop sizes. Based on the theory described above, one would expect a universal behavior that is independent of the drop size and the other problem parameters $c_s$, $H$, $\rho$, and $D$. To demonstrate this, we introduce the nondimensional mass and time as
\begin{equation}
\hat{M}=\frac{M}{\rho R^3}\quad \hat{t}=\frac{c_s-c_\infty}{\rho}\frac{t}{R^2/D}.\label{dimless}
\end{equation}
By substituting (\ref{dimless}) into (\ref{dmdttheory})-(\ref{thetatheory}), we obtain
\begin{eqnarray}\label{dmdtdl}
\frac{d\hat{M}}{d\hat{t}}&=&-\pi f(\theta),\\
\label{mdl}\hat{M}&=&\pi \frac{\cos^3\theta-3\cos\theta+2}{3\sin^3\theta},\\
\label{thetadl}\frac{d\theta}{d\hat{t}}&=&-(1+\cos\theta)^2 f(\theta).
\end{eqnarray}
The relations (\ref{dmdtdl})-(\ref{thetadl}) no longer depend on the size of the droplets, but only on the contact angle. This implies that when we rescale the experimental data according to (\ref{dimless}), they should all collapse onto the theoretical curves described by (\ref{dmdtdl})-(\ref{thetadl}).

\section{Comparison between theory and experiment}\label{comp}
In Sec. \ref{theory} we explained that it should be possible to collapse the experimental data for all droplet sizes measured onto a single theoretical curve. In order to test this, we have to scale the experimental data according to (\ref{dimless}). As a characteristic length scale, we would like to use the droplet radius. However, during the final moments of the droplet's lifetime, the droplet radius is a time-dependent quantity. Therefore, we discarded all data where the droplet radius was changing significantly ($>$10\%) in the results that follow, and used the initial droplet radius for scaling.

The most direct prediction from the Popov model, which involves no time-integration, is the dependence of the rate of mass loss on the contact angle (\ref{dmdtdl}). Indeed, the scaled experimental data collapse onto a single curve, which is in excellent agreement with the theoretical prediction (\ref{dmdtdl}), as shown in Fig. \ref{dmdt}(b). For comparison, the result obtained from applying the model of Hu and Larson \cite{Hu:2002p192} is also shown. Their approximation works well up to $\theta=90^{\circ}$, but for larger contact angles Popov's fully analytical model is required to adequately describe the data.

Figure \ref{modex}(b) shows that the experimental data for the contact angle versus (dimensionless) time follow a universal theoretical curve for all droplet sizes measured.
The total time it takes a droplet to evaporate depends on its initial contact angle, as explained in Sec. \ref{theory}. Since the initial contact angles vary somewhat, the droplet lifetimes differ. However, the experimental time is not an absolute measure, and we therefore have the freedom to set $t=0$ at whichever contact angle we want. As the reference point, we chose $t=0$ at the end of the evaporation process, which is characterized by $\theta=0$. This point is found by linear extrapolation from the last data points measured to $\theta=0$.  

Once the contact angle in time is known, we can apply relation (\ref{mdl}) to derive the droplet mass theoretically. Experimentally, the droplet mass is obtained independently of the contact angle. Therefore, the comparison between the theoretical predictions and the experimental data for the droplet mass, as in Fig. \ref{vol}(b), provides a second validation of the model.

In the results described above, we used the experimental data as long as the contact line remained pinned and hence the droplet radius remained constant. In Fig. \ref{rad}, we showed that depinning occurs during the final moments of the droplet's lifetime. To construct the theoretical curves in Fig. \ref{modex}(a), this radius change has been taken into account. Time integration was performed backwards in time, starting from the smallest contact angle measured. The agreement between the model results and the experimental data is surprisingly good, even in the regime where the droplet radius is changing significantly. Although the droplet radius decreases rapidly, the timescale over which the radius shrinks is still large -in the order of 100 s- compared to diffusion time ($10^{-2}$ s). Therefore, contact-line dynamics is still of negligible influence, and the quasi-steady evaporation model can indeed be applied \cite{Cazabat:2010p1660, Cachile:2002p413}. 

\section{Conclusion}
Evaporation of water droplets on superhydrophobic Carbon Nanofiber substrates is studied. These substrates allowed us to measure the evolution of the droplet mass and contact angle over time, while the contact line remained pinned throughout almost the entire experiment. The initial contact angle was as high as 150$^\circ$, and since it decreases to 0$^\circ$ during evaporation, a very large range of contact angles could be studied. Therefore, CNF substrates are a very useful tool to study droplet evaporation in the absence of contact-line dynamics. In our theoretical analysis, we deduced universal relations for the time-evolutions of the droplet mass and contact angle. This universal scaling behavior is confirmed by our experimental results. Since the experimental data covered almost the entire range of possible contact angles, we have been able to validate the diffusion-based analytical evaporation model presented by Popov \cite{Popov:2005p16}. The agreement of our experimental data with this theoretical model -that does not contain any adjustable parameters- is excellent. Therefore, we conclude that in our experiments the evaporation is quasi-static and diffusion-driven, and thermal effects play no role.  

Even during the brief depinning phase, the quasi-steady model predicted the experimental data surprisingly well. Hence, a pinned contact line is not a stringent requirement for the applicability of the quasi-steady evaporation model, provided that the radius change takes place on a longer timescale than the diffusion. By contrast, for droplets evaporating on complete wetting substrates, a quasi-static droplet profile can no longer be assumed and viscous effects influence the evolution of the contact angle over time \cite{Cachile:2002p413, Poulard:2005p2323, Eggers2010}. It would be interesting to address intermediate cases, in which there is some contact-line motion, so as to establish the range of applicability of the quasi-steady evaporation model. 

\begin{acknowledgments}
Hrudya Nair and Arie van Houselt acknowledge Barbara Mojet for
fruitful discussions.
\end{acknowledgments}

\end{document}